\documentclass[article]{aastex}
\usepackage{emulateapj5}

\def\sqig{$\sim$}
\def\sun{$_\odot$}

\def\cts{counts~s$^{-1}$}
\def\source{GX~13+1}
\def\src{GX~13+1}

\begin{document}
\title{
Properties of the 24 Day Modulation in GX 13+1 from 
Near-Infrared and X-ray Observations}

\author{Robin H.~D. Corbet\altaffilmark{1,2}, 
Aaron B. Pearlman\altaffilmark{3}, 
Michelle Buxton\altaffilmark{4}
\and
Alan M. Levine\altaffilmark{5}}

\altaffiltext{1}{University of Maryland, Baltimore
County, MD, USA; corbet@umbc.edu}

\altaffiltext{2}
{CRESST/Mail Code 662, X-ray Astrophysics Laboratory,
NASA Goddard Space Flight Center, Greenbelt, MD 20771, USA}

\altaffiltext{3}
{Department of Physics, University of Maryland, Baltimore
County, 1000 Hilltop Circle, Baltimore, MD 21250, USA;
aaronp1@umbc.edu}

\altaffiltext{4}
{Department of Astronomy, Yale University, P.O. Box
208101, New Haven, CT 06520, USA}

\altaffiltext{5}
{Kavli Institute for Astrophysics and Space Research, MIT,
Cambridge, MA 02139, USA}

\begin{abstract}
A 24 day period for the low-mass X-ray binary GX 13+1 was previously
proposed on the basis of 7 years of RXTE ASM observations (Corbet
2003) and it was suggested that this was the orbital period of the
system.  This would make it the one of the longest known orbital periods for a
Galactic
low-mass X-ray binary powered by Roche lobe overflow. We present here the results of:
(i) K-band photometry obtained with the SMARTS Consortium CTIO 1.3
telescope on 68 nights over a 10 month interval; (ii) Continued
monitoring with the RXTE ASM, analyzed using a semi-weighted power
spectrum instead of the data filtering technique previously used;
and (iii) Swift BAT hard X-ray observations.
Modulation near 24 days is seen in both the K-band and additional
statistically independent ASM X-ray observations. 
However, the modulation in the ASM is not
strictly periodic. 
The periodicity is also not detected in the Swift BAT
observations, but modulation at the same relative
level as seen with the ASM cannot be ruled out.
If the 24 day period is the orbital period of system, this implies
that the X-ray modulation is caused by structure that is
not fixed in location. 
A possible mechanism for the X-ray modulation
is the dipping behavior recently reported from
XMM-Newton observations.

\end{abstract}

\section{Introduction}

GX 13+1 is a bright low-mass X-ray binary (e.g. Homan et al. 2004
and references therein)
that has 
rarely exhibited X-ray bursts (Fleischman 1985; Matsuba et al. 1995),
showing that the compact object in the system is a neutron star. 
An infrared counterpart was identified by
Naylor et al. (1991) and Garcia et al. (1992).  The infrared counterpart was
previously found to vary on timescales of days to tens of days, although no
definite orbital period was detected (e.g., Charles \& Naylor 1992; Groot
et al. 1996; Bandyopadhyay et al. 2002). From infrared spectroscopy
Bandyopadhyay et al. (1999) derived a spectral type of K5 III for the
mass-donating star. This classification implies a mass of 5
M\sun (Allen 1973), and so the mass donor is the primary
star, unlike the majority of low-mass X-ray binaries (LMXBs).

A search was previously made (Corbet 2003; hereafter C03) 
for periodic modulation in
the X-ray flux from GX 13+1 using Rossi X-ray Timing
Explorer (RXTE) All-Sky Monitor (ASM) data collected over an
interval of almost 7 years. From a filtered data set, which excluded
measurements with large uncertainties, modulation was found at a period of
24.065$\pm$0.018 days. The modulation was most clearly detectable at high
energies (5-12 keV). Spectral changes were revealed as
a modulation in hardness ratio on the 24 day period, and there was a
phase shift between the modulation in the 5 - 12 keV
energy band and the 1.5 - 5 keV band. The high-energy
spectrum of GX 13+1 is unusual in displaying both iron emission and
absorption line features, and it was speculated in
C03 that the peculiar
spectral and timing properties may be connected.
Bandyopadhyay et al. (2002) proposed that the 24 day modulation
was part of a long timescale quasi-periodic modulation.

Because of the unusual nature of this modulation, and the somewhat
nonstandard technique used to maximize the signal in the ASM data, it
was desired to confirm the 24 day period through further observations
and determine whether the modulation is present at other wavelengths.
We present here additional both additional RXTE ASM data
and also K-band observations
that confirm the presence of the 24 day period. However, the 24 day
period is found not to be strictly periodic in the ASM data. We
also analyze Swift Burst Alert Telescope (BAT) observations 
which do not show significant modulation near 24 days.
We discuss possible causes of the modulation and suggest that
it may be caused by
dipping.

\section{Observations}
\subsection{X-ray: RXTE ASM}
The RXTE ASM (Levine et al. 1996) consists of three similar
Scanning Shadow Cameras (SSCs), sensitive to X-rays in an energy band of
approximately 1.5-12 keV, which perform sets of 90 second pointed
observations (``dwells'') so as to cover 
as much as \sqig80\% of the sky every
\sqig90 minutes.  
Light curves are available in three energy bands: 1.5 to 3.0 keV, 3.0
to 5 keV, and 5 to 12 keV.
The Crab produces approximately 75
\cts\ in the ASM over the entire energy range. Observations
of blank field regions away from the Galactic center indicate that
background subtraction may produce a systematic uncertainty of about 0.1
\cts\ (Remillard \& Levine 1997). 
The ASM light curve of \src\ considered here covers
approximately 14 years (MJD 50,088 to 55,267; 1996-01-06 to 2010-03-12).
For the reasons discussed in Section 3.1 we use ASM light
curves binned to one day time resolution.  

\begin{figure}
\epsscale{0.9}
\includegraphics[angle=-90,width=7.5in]{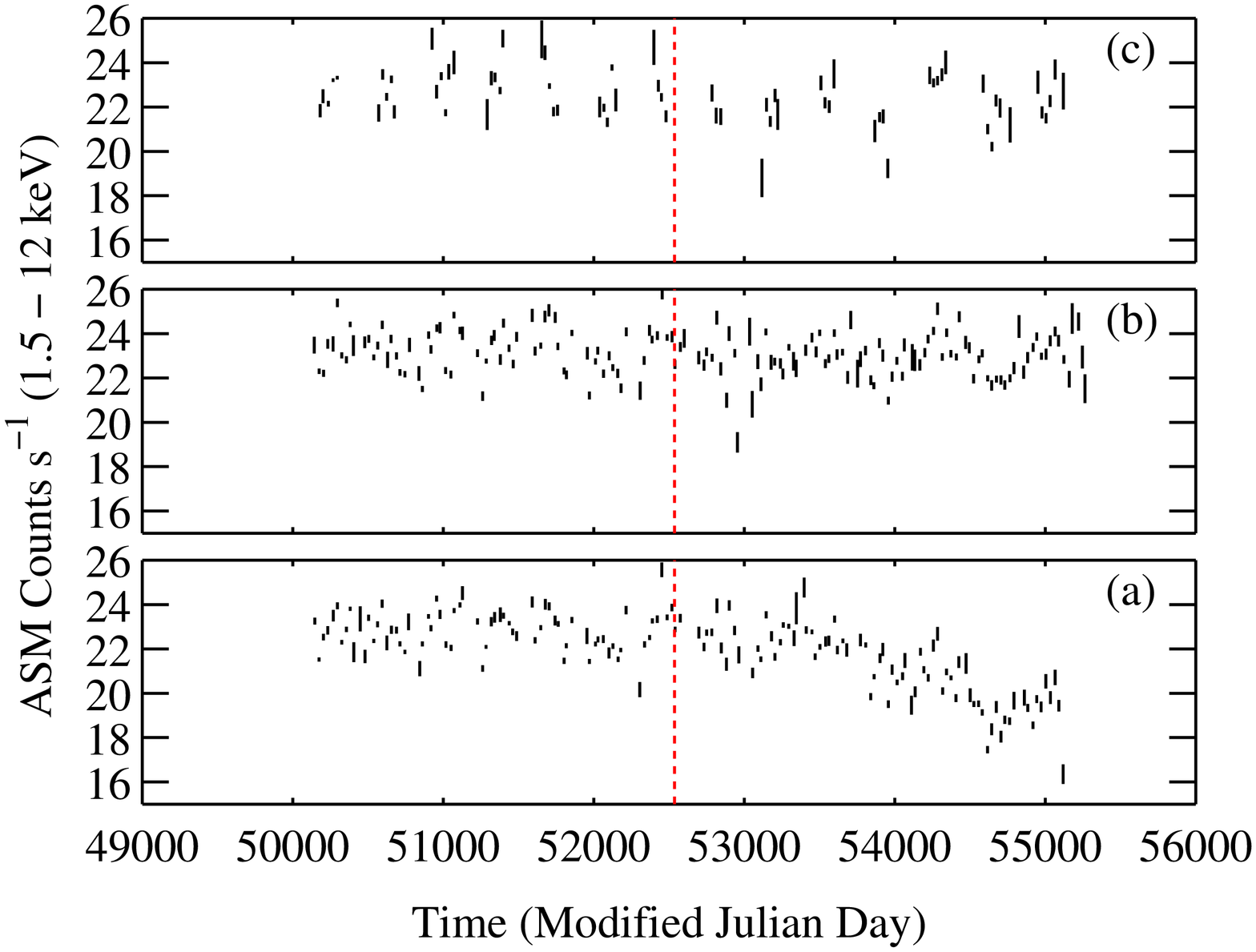}
\figcaption[f1.eps]{
The RXTE ASM light curve of \src\ in 30 day averages.
Only time bins which contain a minimum of 20 dwells
are plotted.
The three panels show data from each Scanning Shadow
Camera separately: (a) SSC 1; (b) SSC 2; (c) SSC 3.
The vertical dashed lines indicate the end of the time range
used in C03.
SSC 1 shows a flux decline in the second half
of the light curve related to gain changes
in this detector.
\label{fig:ssc_lc}
}
\end{figure}

\begin{figure}
\epsscale{0.9}
\includegraphics[angle=-90,width=7.5in]{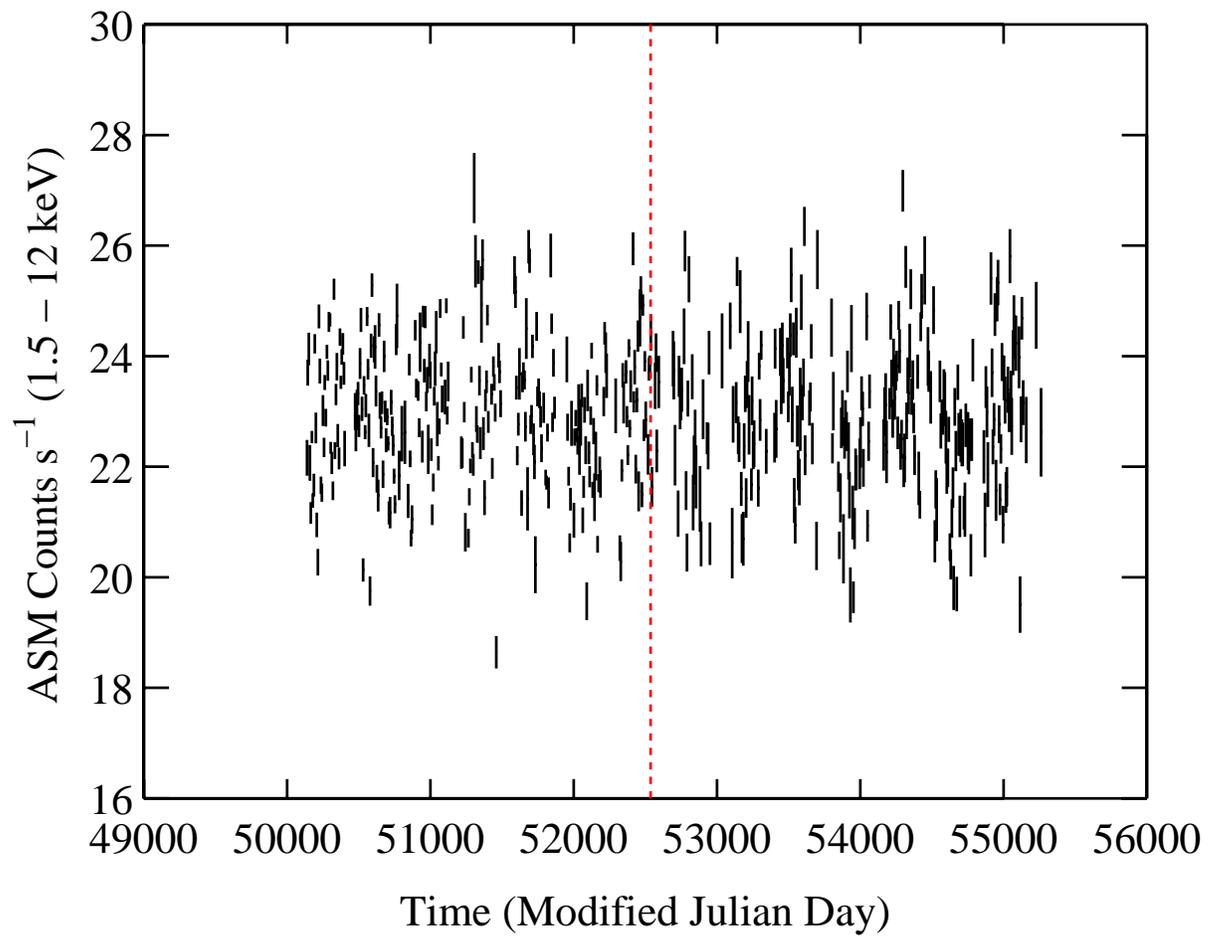}
\figcaption[f2.eps]{
The RXTE ASM light curve of \src\ in 1 week averages.
Only time bins which contain a minimum of 20 dwells
are plotted.
The vertical dashed line indicates the end of the time range
used in C03.
Only data from SSCs 2 and 3 are included for times after
that indicated by the dashed line.
\label{fig:asm_lc}
}
\end{figure}

The three SSCs (``1'', ``2'', and ``3'') that make up the ASM have experienced 
changes in response with SSC 1 having experienced 
a gain change of about 10\% per year
(Levine et al. 2010).
We therefore investigated the ASM light curve of \src\
considering the three SSCs separately.
The light curves obtained in this way are shown in
Figure \ref{fig:ssc_lc}. It is seen that although
similar count rates are obtained from the three detectors
for approximately the first half of the light
curve, during the second half an apparent decline in flux
occurs only in the light curve from SSC 1.
This effect is likely to be due to instrumental changes
and we therefore only use SSC 1 data for times before
MJD 52,536, i.e. the data range used in C03.
The overall RXTE ASM light curve
of \src\ obtained with this detector selection is shown in 
Figure \ref{fig:asm_lc}. No long-term trend
is obvious and the mean
flux for the entire ASM energy range is 22.90 $\pm$ 0.01
(statistical)  \cts.
We also investigated the light curve in the three available
energy bands for each SSC. It was found that the energy-separated
light curves of SSC 1 show systematic differences from the light curves
in the other two SSCs. For this reason we only use
data from SSC 2 and 3 when examining energy-separated ASM light curves.

\subsection{Infrared: SMARTS/ANDICAM}
\begin{figure}
\epsscale{0.9}
\includegraphics[angle=-90,width=7.5in]{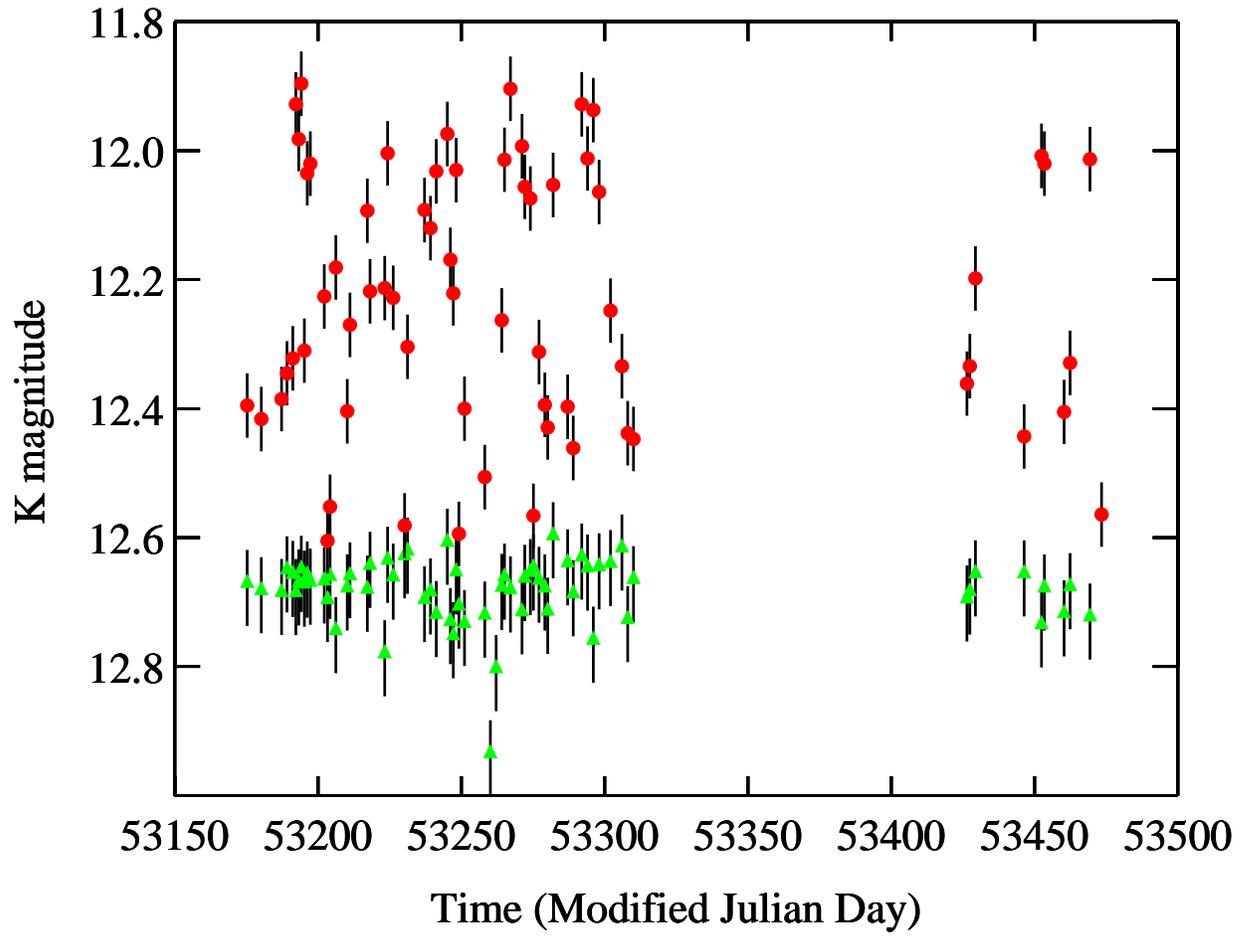}
\figcaption[f3.eps]{
SMARTS CTIO ANDICAM K-band photometry of \src\ (red filled circles) 
and the comparison star NCL 107 (green filled
triangles).
\label{fig:ir_lc}
}
\end{figure}

We used the SMARTS Consortium CTIO 1.3m telescope and ANDICAM detector
(DePoy 2003) to obtain 340 infrared images in the K-band ($\lambda$ 
= 2.2 $\mu$m). There were 5 images taken per day for a total of 68 days over a
10 month interval. The observations span the period from MJD 53,175
(2004-06-19) to 53,473 (2005-04-13).
We performed profile-fitting photometry using DAOPHOT after flat
fielding the original images with an image constructed from the
difference of dome flats obtained with the flat field lamps on and off
(Bandyopadhyay et al. 2002). Each set of 5 infrared images was
shifted, aligned, and combined into a final image that was suitable
for performing relative photometry.
We employed the same star,
number 103 of Naylor et al. (1991; ``NCL'') for all fields
to define the profile for each night for the fits. Photometric measurements
were also all made relative to NCL 103.
Conversion to absolute magnitude was made using the mean
flux
of NCL 106 and the value of K = 12.69 given in Naylor et al. (1991).
The standard deviation of the brightness of Star NCL 106 is
0.05 mag. and we adopt this as an approximation of the
uncertainty of the measurements of \src.
The resulting light curve is plotted in Figure \ref{fig:ir_lc}
and
\src\ is clearly seen to be highly variable.

\subsection{X-ray: Swift BAT}

The Swift BAT is described in detail by Barthelmy et al. (2005).
It is a wide field-of-view instrument that comprises a coded mask 
aperture with a CdZnTe detector.
The data used here comes from the Swift/BAT transient monitor
results provided by the Swift/BAT team. This provides a light
curve covering the energy range 15 to 50 keV.
The light curve used here spans the time range
of MJD 53,414 (2005-02-13) to 55,267 (2010-03-12)
and we rebinned the provided orbital light curve
to 1.0 day resolution for our analysis for the reasons discussed
in Section 3.1.

\section{Analysis and Results}

\subsection{Power Spectrum Weighting}
RXTE ASM light curves comprise measurements with 
a very wide range of error bar sizes. 
This means that the error should be taken into account
in the calculation of the contribution of each data point
to a power spectrum.
Scargle
(1989) proposed that the effect of unequally weighted data points could
be understood by considering the combination of points that coincide.
The weighting of a power spectrum is thus analogous to
the calculation of a weighted mean. However, as
with the weighted mean, in practice the choice of weighting factors
needs to be made carefully. 
	
In C03 it was argued that, due to the large flux variations
compared to the error bar size, simply weighting data points
by just the size of their errors was not
appropriate and actually decreased the sensitivity of the power
spectrum for period detection. A simple data filtering technique was
used instead: points with large error bars were excluded, with the
threshold chosen to maximize the strength of the 24 day
modulation. Although this gave apparently good results, a drawback is that the
choice of filtering threshold depends on the assumption that the peak
being maximized is indeed a real signal.

In Corbet et al. (2007a) a modified weighting scheme was proposed that
could deal with any degree of flux variability compared to error bar
size. This was later noted (Corbet et al. 2007b) to be equivalent to
the semi-weighted mean (Cochran 1937, 1954). This technique makes no a
priori assumption about the presence of a periodic signal in a data set. The
weight chosen for each data point is a combination of both the error
bar on that point and the estimated source variability.
The source variability is determined by calculating the
excess variance of the light curve compared to
that predicted from the uncertainties of the data points. 
Because the semi-weighting scheme does not make
any assumptions about the presence of a periodic signal
in the data set, and gives appropriate weighting for sources
of any brightness, we 
therefore use semi-weighting in this paper. 
In addition, because the number of ASM and BAT observations
per day varies due to orbital precession and other effects
for both instruments, we rebinned the light curves to
one day time resolution to avoid
the over-weighting of times with larger numbers
of dwells
that would be effectively introduced otherwise (see e.g. C03).
This over-weighting adversely affects power spectra and may
introduce artifacts, even if
semi-weighting is used.
The infrared photometry data have approximately equal error bars and there
is thus no benefit to weighting these data.
For all the power spectra presented in this paper we oversample
by a factor of 3 compared to the nominal frequency resolution
of each power spectrum.

\subsection{RXTE ASM}

\begin{figure}
\epsscale{0.9}
\plotone{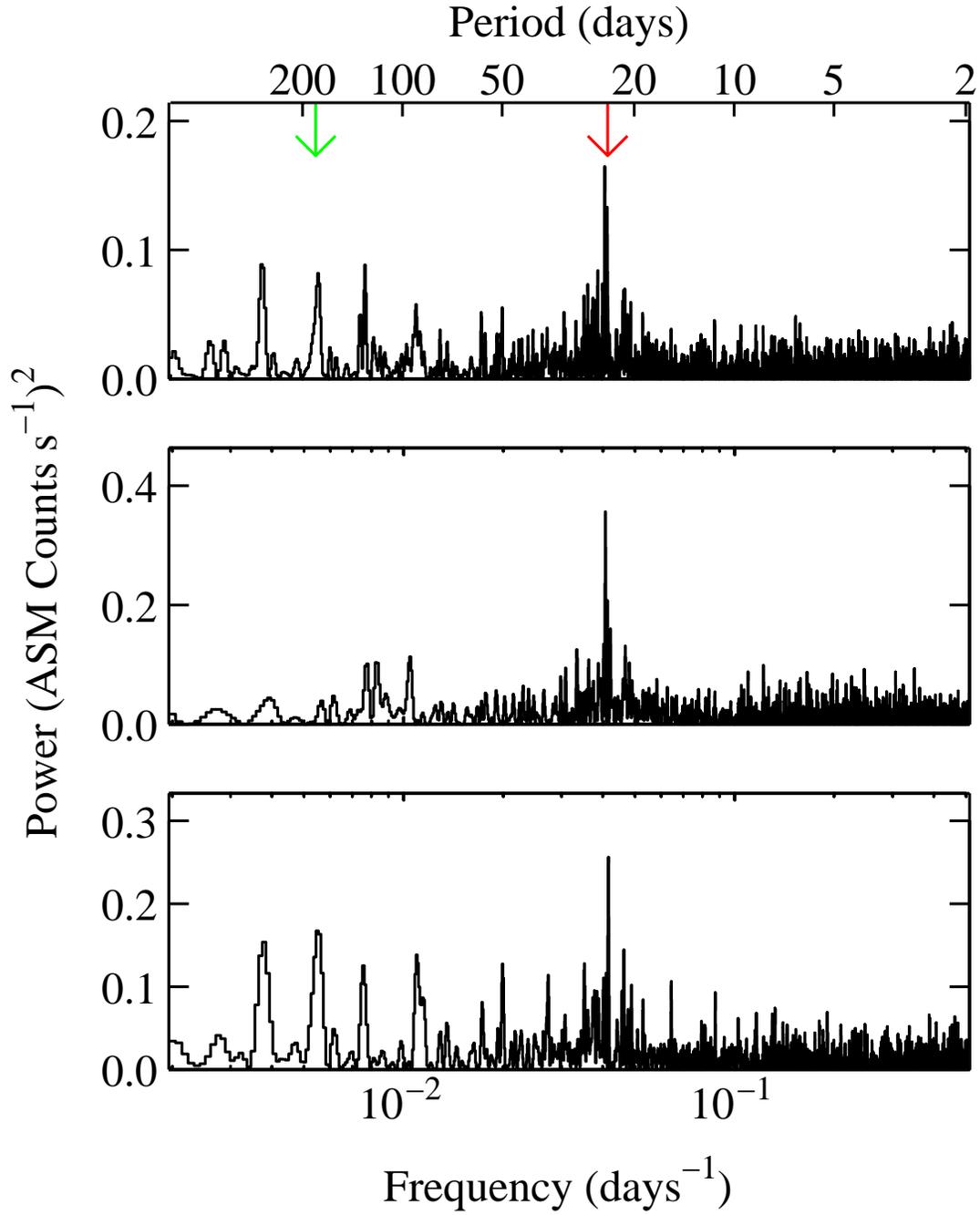}
\figcaption[f4.eps]{
Power spectra of the RXTE ASM observations of \src.
The red arrow marks the period reported in C03
and
the green arrow marks half a year which is a common
artifact in power spectra of ASM light curves.
Bottom panel: power spectrum of data presented in C03;
middle panel: power spectrum of data obtained since C03;
top panel: power spectrum of all data.
\label{fig:asm_power}
}
\end{figure}

We calculated semi-weighted power spectra of the RXTE ASM light curve
using three different data selections: (i) the same data used in
C03, \sqig6.7 year duration; (ii) only data obtained since
C03, \sqig7.5 year duration; and (iii) the entire ASM light
curve, \sqig14.2 year duration. These 
power spectra are shown in Figure \ref{fig:asm_power}.

\begin{figure}
\epsscale{0.9}
\includegraphics[angle=-90,width=7.5in]{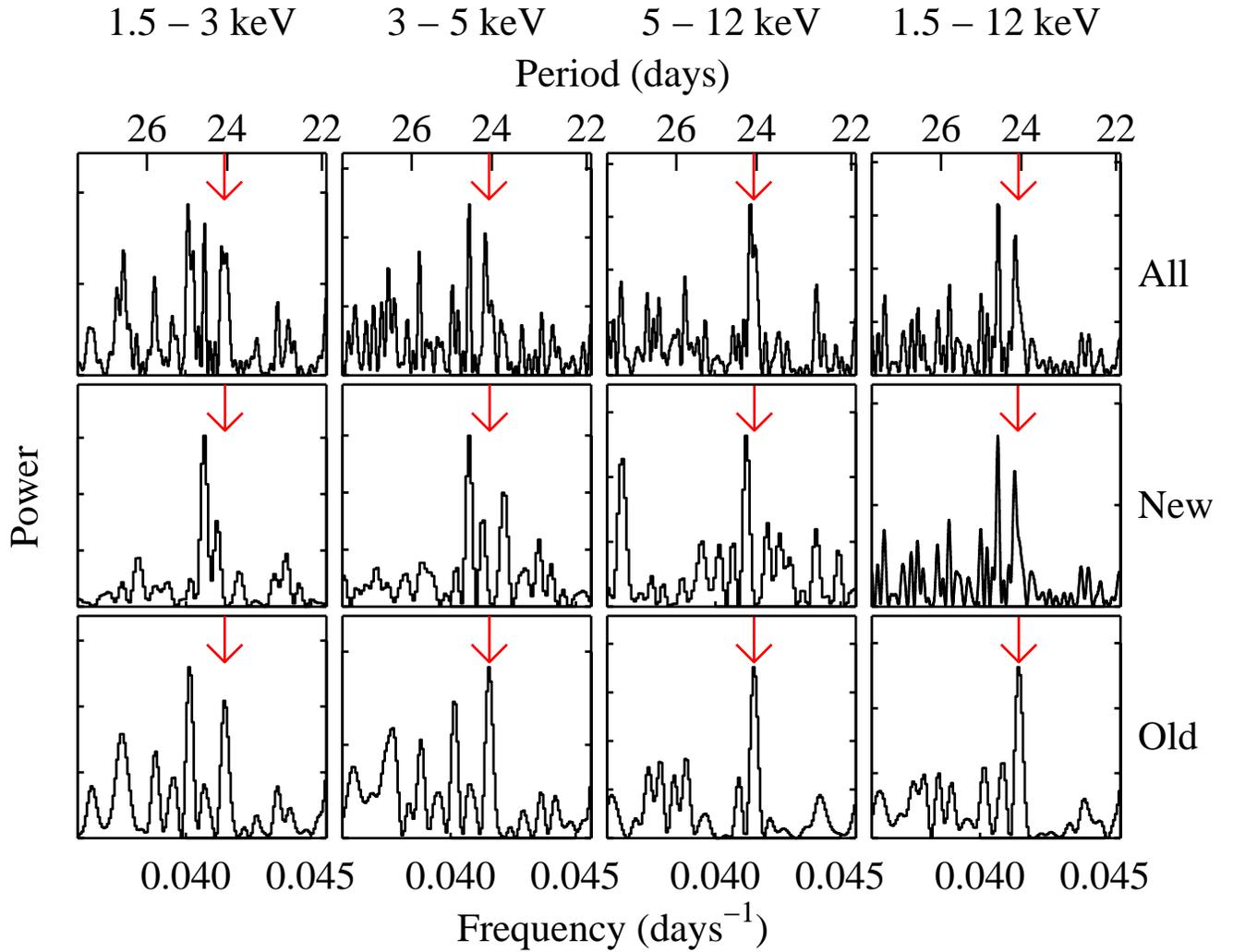}
\figcaption[f5.eps]{
Power spectra of RXTE ASM observations of \src\
separated by time intervals and energy band.
Bottom panels: power spectra of data presented in C03;
middle panels: power spectra of data obtained since C03;
top panels: power spectra of all data.
The arrows mark the period reported in C03.
Note that the power spectra are oversampled by a factor
of 3 compared to their nominal resolution.
\label{fig:asm_old_new_band}
}
\end{figure}

In data set (i) a peak is again found at the proposed 24 day
period. In data set (ii), which is statistically independent, the
strongest peak in the 2 to 500 day period range is also found near
24 days. In the full data set (iii) the 24 day period is the strongest
feature in the power spectrum.  However, when the peak locations are
examined (Fig. \ref{fig:asm_old_new_band}), 
it can be seen that the peak location is not constant
 - either between time ranges or among
the different energy ranges.
It is common to estimate signal coherence using
a quality factor, ``Q'', defined as the frequency
of a modulation divided by its width in the power spectrum
(e.g. van der Klis 2000). However, this simple characterization
does not lend itself well to the modulation seen with the ASM
in \src\ where, rather than just a broad peak, we see multiple sharp
peaks.

\begin{figure}
\epsscale{0.9}
\plotone{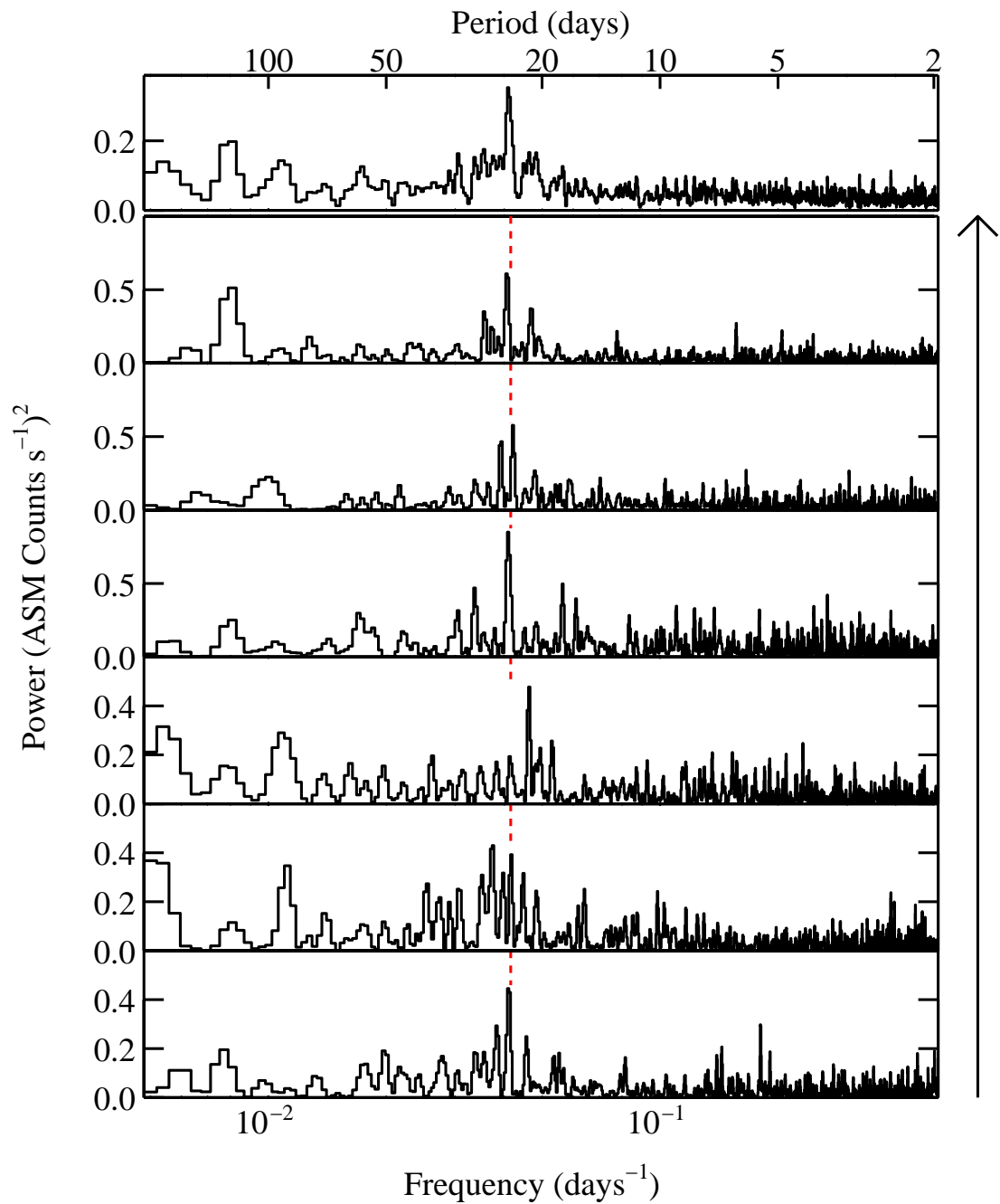}
\figcaption[f6.eps]{
Power spectra of RXTE ASM observations of \src\
divided into 6 equal time intervals. The top panel
shows a non-coherent sum of the power spectra in the 6
lower panels.
The dashed red lines mark the period reported in C03.
The arrow to the right of the figure indicates increasing time.
\label{fig:multi_asm_ft}
}
\end{figure}

We next examined the 1.5 to 12 keV ASM light curve by dividing
up the light curve into several equal length sections and taking the power
spectrum of each of the sections. 
Similar results were obtained for a wide range of number of
sections used to divide the light curve,
and the results using 6 sections are plotted in
Figure \ref{fig:multi_asm_ft}. It can be seen that in
several of the light curve sections a peak is present near
24 days, but it is not constant in strength.
In the top panel of Figure \ref{fig:multi_asm_ft} we also
show a simple sum of the power spectra of the
individual light curve sections (i.e. signal coherence is
not considered in adding the power spectra). This shows a prominent
peak centered on approximately 24 days.
We fitted a Gaussian function to the peak in
the light curve and obtained a period
of 24.27 days with a formal statistical error of
0.03 days.

\subsection{Infrared}
\begin{figure}
\epsscale{0.9}
\includegraphics[angle=-90,width=7.5in]{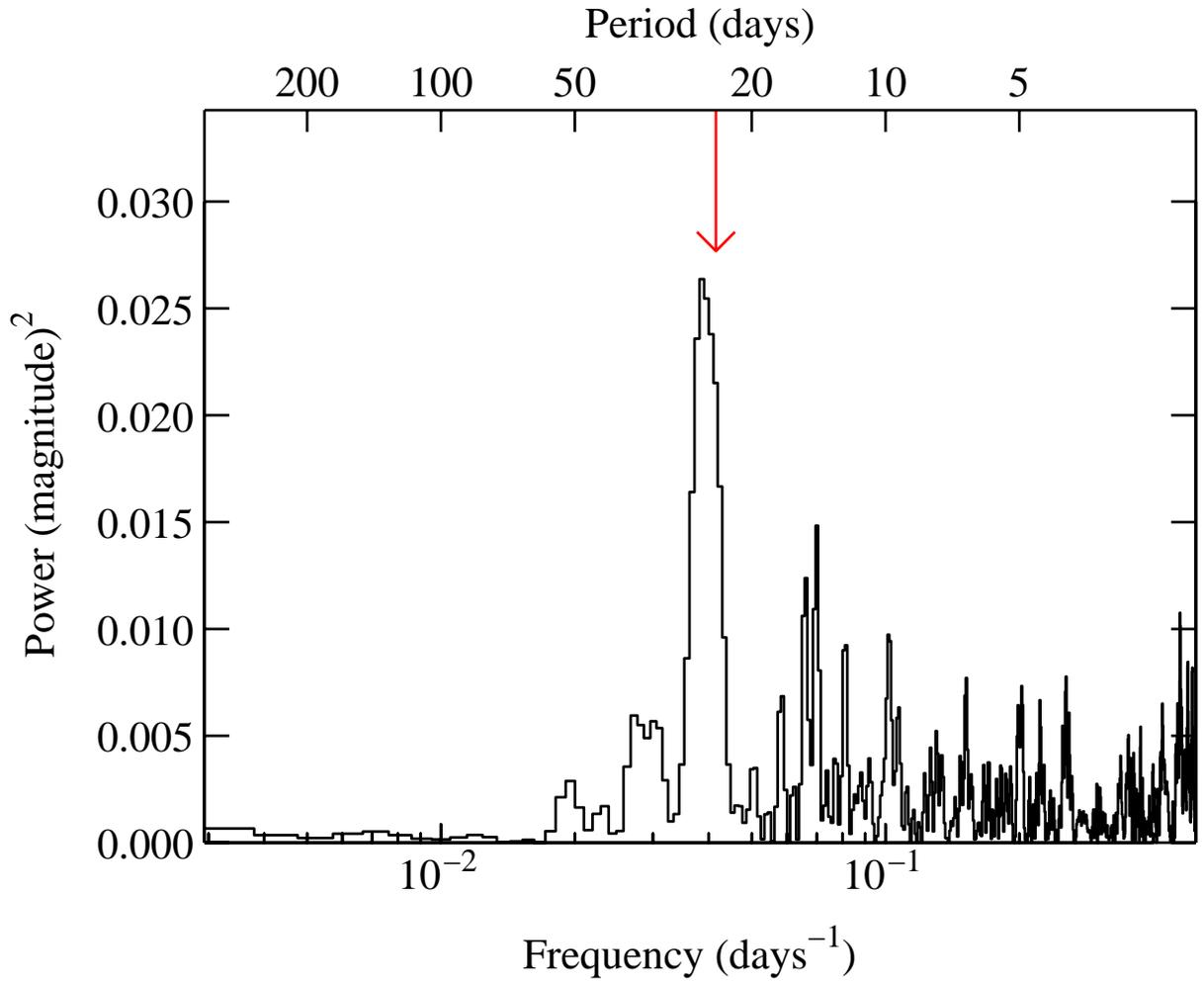}
\figcaption[f7.eps]{
Power spectrum of the SMARTS CTIO ANDICAM K-band photometry of \src.
The red arrow marks the period reported in C03.
\label{fig:ir_power}
}
\end{figure}

The K-band light curve of \src\ (Fig. \ref{fig:ir_lc}) clearly shows variability similar to that seen
before in shorter observations. We calculated an unweighted
power spectrum of the light curve to search for periodic modulation
(Fig. \ref{fig:ir_power}) and this
shows a strong peak near 24 days.
The false alarm probability (FAP; Scargle 1982) for finding a peak of this
strength anywhere in the {\em entire} period range searched of
2 to 298 days is 0.5\%. The estimated FAP is likely
to be somewhat of an underestimate, because the light curve is likely to contain,
in addition to periodic modulation, non-periodic components of unknown noise properties.
This is seen in optical observations of several LMXBs (e.g. Corbet et al. 1986, 1989;
Harris et al. 2009).
From fitting
a sine wave to the light curve we obtain a period of 25.8 $\pm$ 0.3
days 
which is similar to, although formally not overlapping,
the 24.065$\pm$0.018 days period reported
in C03.

We investigated whether it was possible to combine our infrared
observations with the light curves published by Charles \&
Naylor (1992) and Bandyopadhyay et al. (2002) to refine the period
measurement. However, the very large intervals between the
observations resulted in severe aliasing in the power spectrum
which made it impossible to improve the precision of the period
determination.

\subsection{Swift BAT}
\begin{figure}
\epsscale{0.9}
\includegraphics[angle=-90,width=7.5in]{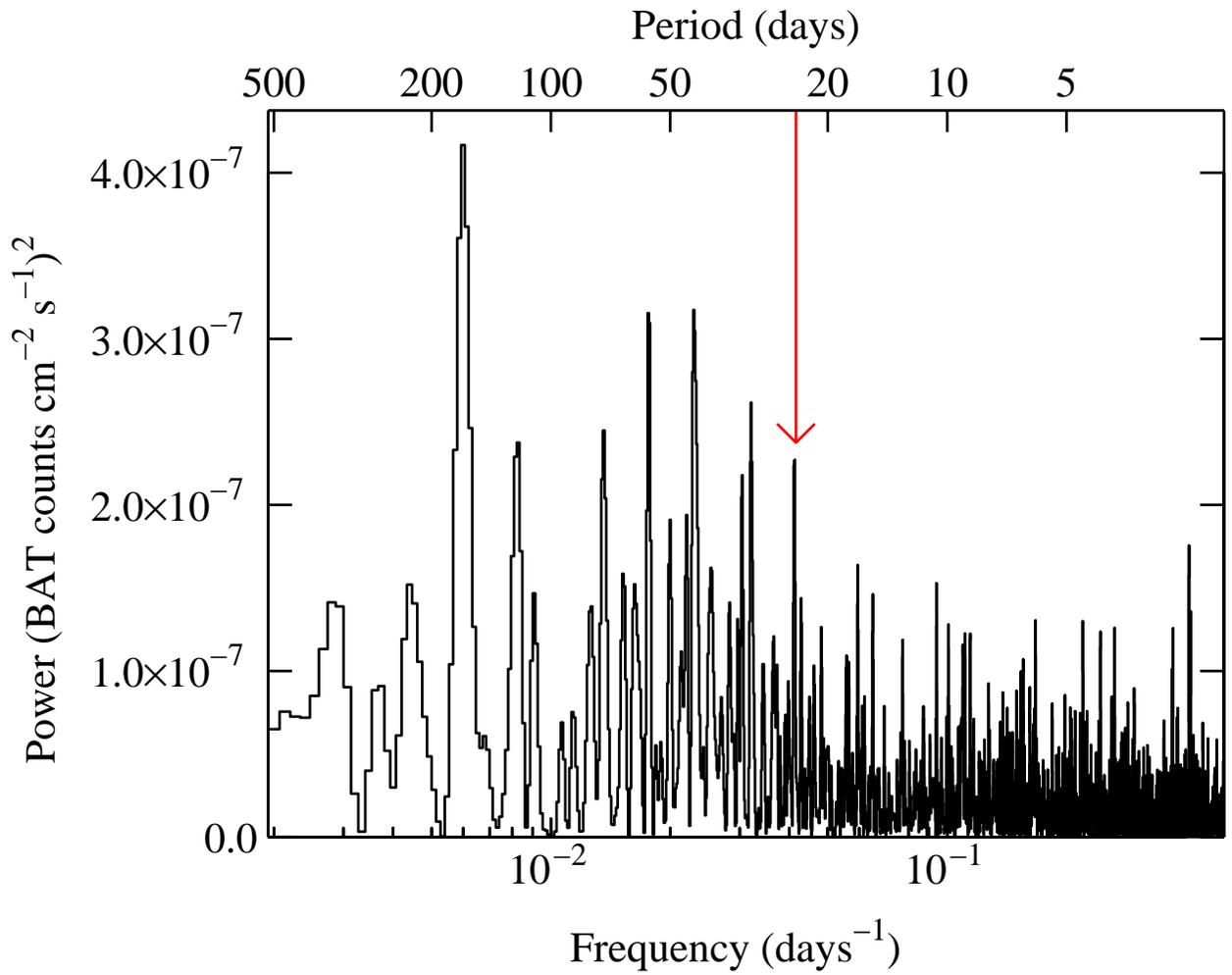}
\figcaption[f9.eps]{
Power spectrum of the Swift BAT light curve
of \src. The red arrow indicates the period
reported in C03.
\label{fig:bat_power}
}
\end{figure}

A power spectrum of the semi-weighted Swift BAT light curve does
not show strong modulation at the 24 day period
(Fig. \ref{fig:bat_power}), although there is a small
peak of very low significance near the previously reported
24 day period.
If we take the small peak near 24 days in the power spectrum of the 
BAT light curve of \src\ to be a detection, then
the modulation in this energy band (Fourier amplitude divided by
mean flux) would be 
9\%.
For comparison, the
``incoherent'' power spectrum of the ASM
full energy band light curve (Fig. \ref{fig:multi_asm_ft}),
has an amplitude (Fourier amplitude/mean flux) of
2.4\%. 
Thus, the non-detection of modulation with the BAT
does not exclude the presence of
modulation at the level seen with the ASM.

\section{Discussion}
\begin{figure}
\epsscale{0.9}
\includegraphics[angle=-90,width=7.5in]{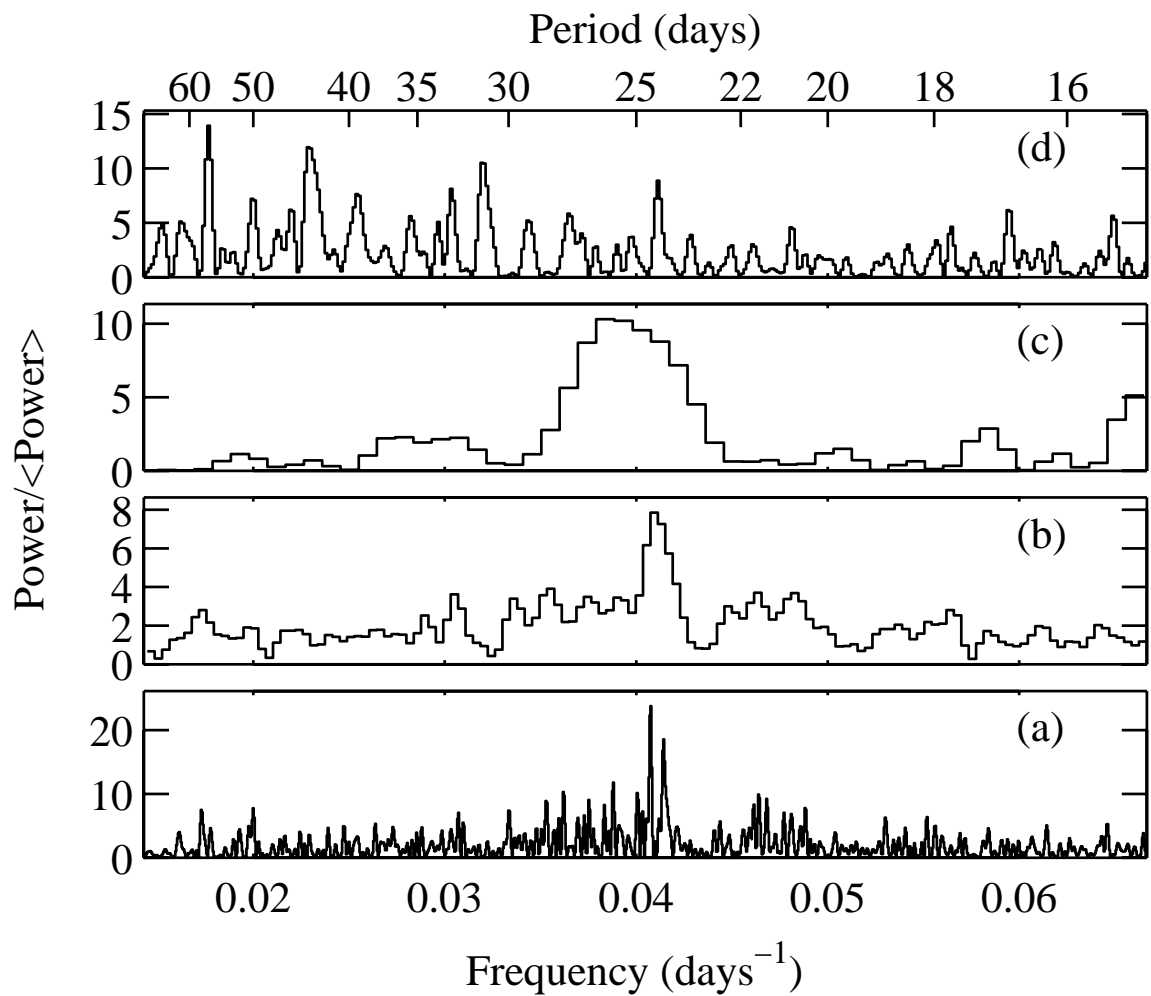}
\figcaption[f10.eps]{
Comparison of the power spectra of \src\ from
(a) RXTE ASM (power spectrum of the entire
light curve); (b) RXTE ASM (non-coherent sum of power spectra
of the separate sections of the ASM light curve shown
in Fig. \ref{fig:multi_asm_ft});
(c) K-band photometry; (d) Swift BAT.
In all cases the power is normalized by the mean power
measured over a frequency range of 1/200 to 1/2 days$^{-1}$. 
\label{fig:mega_power}
}
\end{figure}

The previously proposed 24 day period for \src\ is
seen in subsequently obtained,
statistically independent, ASM data and is also
present in the infrared light curve.
The period is also detected in a
re-analysis of the initial ASM light curve using a semi-weighted power
spectrum.
Although the periodicity is not clearly
seen in the Swift BAT light curve,
the BAT observations appear to have lower sensitivity to
a given fractional modulation in \src\ than the ASM observations.
A ``summary'' plot comparing the various power spectra is shown
in Figure \ref{fig:mega_power}.
The multiple peaks in the power spectrum of the ASM light
curve (Fig. \ref{fig:multi_asm_ft}) indicates that
the X-ray modulation is not strictly periodic and
it is therefore difficult to compare the relative phasing of the
X-ray and infrared modulations. If a power
spectrum of the ASM light curve is calculated using only
data obtained during the interval
when the infrared observations were made, then no modulation is seen,
presumably because the amount of data included does not give
sufficient sensitivity to
see the modulation. 
However, a significantly longer stretch of
X-ray data cannot be trivially folded: the uncertainty of the
period derived from the infrared data alone is too large, and the
X-ray data do not provide a unique period to use
for folding.

Given the weakness of the X-ray modulation and its lack
of coherence it must be considered whether this could be an
artifact. However, modulation near 24 days is
not seen in other sources with the ASM (e.g.
Farrell et al. 2005, Wen et al. 2006).
In addition, the modulation seen near this period in
the infrared strongly supports an astrophysical rather than an instrumental
origin.

In principle, the 24 day modulation could be
caused by either an orbital period or a super-orbital
period. We note that this period is of the length expected for the
orbital period of a Roche-lobe filling K5 III star 
(Bandyopadhyay et al. 1999). Even though there are uncertainties
in the spectral classification, and the radius of the star may differ
from that expected from its spectral type due to evolutionary effects
caused by mass transfer in the binary, it would be surprising for
the expected orbital period to be hugely different from 24 days.
In Roche-lobe filling systems with super-orbital modulation such
as Her X-1, LMC X-1, and SMC X-1, the super-orbital periods are about
an order of magnitude longer than the orbital periods (e.g. Charles et
al. 2008). 
In addition, the detection of infrared modulation on only the 24 day
period and no other period suggests an orbital origin.
An orbital
interpretation of the 24 day period is therefore favored for \src. 
This orbital period would be the one of the longest
known for a Galactic LMXB accreting via Roche-lobe overflow.

Modulation on the orbital period in the optical and infrared has been seen for a
number of LMXBs (e.g. Charles \& Coe 2003). Depending on the X-ray
luminosity, the waveband observed, and the system inclination, key
contributions to modulation can come from ellipsoidal modulation of
the mass-donor (which would yield two maxima and minima per orbit),
the varying aspect of the heated face of the donor, and eclipses of
the accretion disk.
Periodic modulation of the X-ray flux of LMXBs on the orbital
period is much less commonly observed (e.g. White \& Mason 1985). In systems with high
inclination angles eclipses may occur. However, in such systems the
central X-ray source is generally obscured by the rim of the accretion
disk and X-rays are only observed which are scattered from an
accretion disk corona (ADC).
At somewhat lower inclination angles, systems might exhibit
dips that are thought to be caused by structure at the rim of
the accretion disk.
Dipping in X-ray binaries is reviewed by, for example, White
\& Mason (1985) and Frank et al. (1987).

The dipping behavior reported for \src\ by D\'{\i}az Trigo et al. (2010)
from XMM-Newton observations suggests a possible mechanism for
the X-ray modulation in this source. Since the dipping would be expected to
be variable in depth and location, this would provide a natural
explanation for the lack of strict periodicity found
in the ASM light curves. 
If this dipping interpretation is correct, the location
of the material causing the dips could vary and thereby
tend to broaden any sharp peaks in a power spectrum.
Lower energy X-rays might also be more
strongly attenuated by photoelectric absorption
leading to a larger orbital modulation.
The ASM light curve is only modulated by a few percent on average
over the 24 day period and thus, a modest degree of dipping
could explain the X-ray modulation.
If dipping is indeed present, then it would be expected
that this should cause modulation on the orbital period.
If the 24 day period is not the orbital period of \src\
then the lack of a detection of the orbital period 
in the power spectrum would be puzzling.
A possible peculiarity, however, is that although the 
modulation near 24 days is not strictly periodic
in X-rays, we do not see any evidence for harmonics of 
the modulation in the power spectrum. Typically in
dipping systems the modulation is non-sinusoidal and
so harmonics might be expected to be seen.
Another possibility is that structure at the edge of the disk
modulates only the portion of the X-rays that comes from
an ADC. This might result in modulation that
is closer to sinusoidal. However, if the modulation reflects
an underlying orbital period, the lack of strict periodicity
in an ADC interpretation would be surprising. For example,
the ASM light curves of the ADC sources X\,1822-371
and X\,2127+119 do not show any evidence for period
changes (Wen et al. 2006).
The infrared modulation would be expected to be caused
by one or more of the mechanisms described above (ellipsoidal variations,
X-ray heating of the donor, accretion disk eclipses) none of which
are as susceptible to modulation phase changes as modulation in
the X-ray band such as dipping which is caused by structure in the outer
accretion disk.

It is
potentially instructive to compare the variability of \src\
with other LMXBs with long orbital periods.
The well-studied LMXB  
Cygnus X-2 has a relatively long orbital period of 9.8 days 
(Casares et al. 1998, Elebert et al. 2009). 
For Cyg X-2, Orosz \& Kuulkers (1999) found
that the B and V light curves, when folded
on the orbital period, are dominated by ellipsoidal variations
and that X-ray heating of the mass-donor is relatively 
unimportant.  
Orosz \& Kuulkers (1999) suggest that X-ray heating may be unimportant
for Cyg X-2 due to a large orbital separation and
because a thick accretion disc shields the surface of
the mass-donor star. 
However, in the case of \src\ it seems unlikely that the orbital
period is twice the 24 day period as the mass-donor would then
not fill its Roche-lobe, unless the radius of the mass-donor
is indeed much different from that predicted from its apparent
spectral class. Thus, the infrared modulation in
\src\ may be more likely to be caused by X-ray heating of
the mass-donating star.
For the Galactic black hole candidate systems
1E 1740.7-2942 and GRS 1758-258, periods
of 12.7 and 18.5 days respectively have been proposed
by Smith et al. (2002). The proposed modulation fractions
are 3-4\% for both sources and Smith et al. (2002)
suggest that the modulations are orbital in origin and that
the mass-donors are both red giants. These two sources
may thus have some similarities with \src, even though
the accreting objects are likely to be black holes rather
than neutron stars.
GRS 1915+105 has an orbital period of 30.8 days (Neil
et al. 2007), comparable to the period of \src. 
However GRS 1915+105 is rather different from \src\
because it is a microquasar system containing a massive
black hole (e.g. Greiner et al. 2001).
Long orbital periods have also been reported for
extragalactic systems such as the source in M82, which has
a 62 day period (Kaaret \& Feng 2007), and NGC 5408 X-1,
which has a 115 day period (Strohmayer 2009). However, both these
systems are ultraluminous X-ray sources, probably containing
black holes, and so may well also be quite different 
from \src.

\section{Conclusion}

A \sqig24 day period close to that previously proposed for GX 13+1 
(C03) is seen in both
new ASM data and near-infrared observations. 
However, the X-ray modulation is not strictly coherent
which suggests that it may be caused by structure that is
not completely phase-locked in the binary system.
We propose that the X-ray variability may be caused by
dipping behavior.
A valuable contribution to an
understanding of the system
would be a radial velocity orbit of the mass-donor and/or
the accretion disc from infrared
spectroscopy which could show whether the 24 day period is
indeed the orbital period, and determine the mass function
and also the location of the system components
as a function of phase. Additional infrared photometry has the potential
to more accurately determine the orbital period.

\acknowledgments
We thank Tim Naylor for providing us with the photometric data from 
Charles \& Naylor (1992). We thank Luigi Stella and an anonymous referee
for useful comments. This paper made use of Swift/BAT transient
monitor results provided by the Swift/BAT team.

\pagebreak

\begin{figure}
\epsscale{0.9}
\includegraphics[angle=-90,width=7.5in]{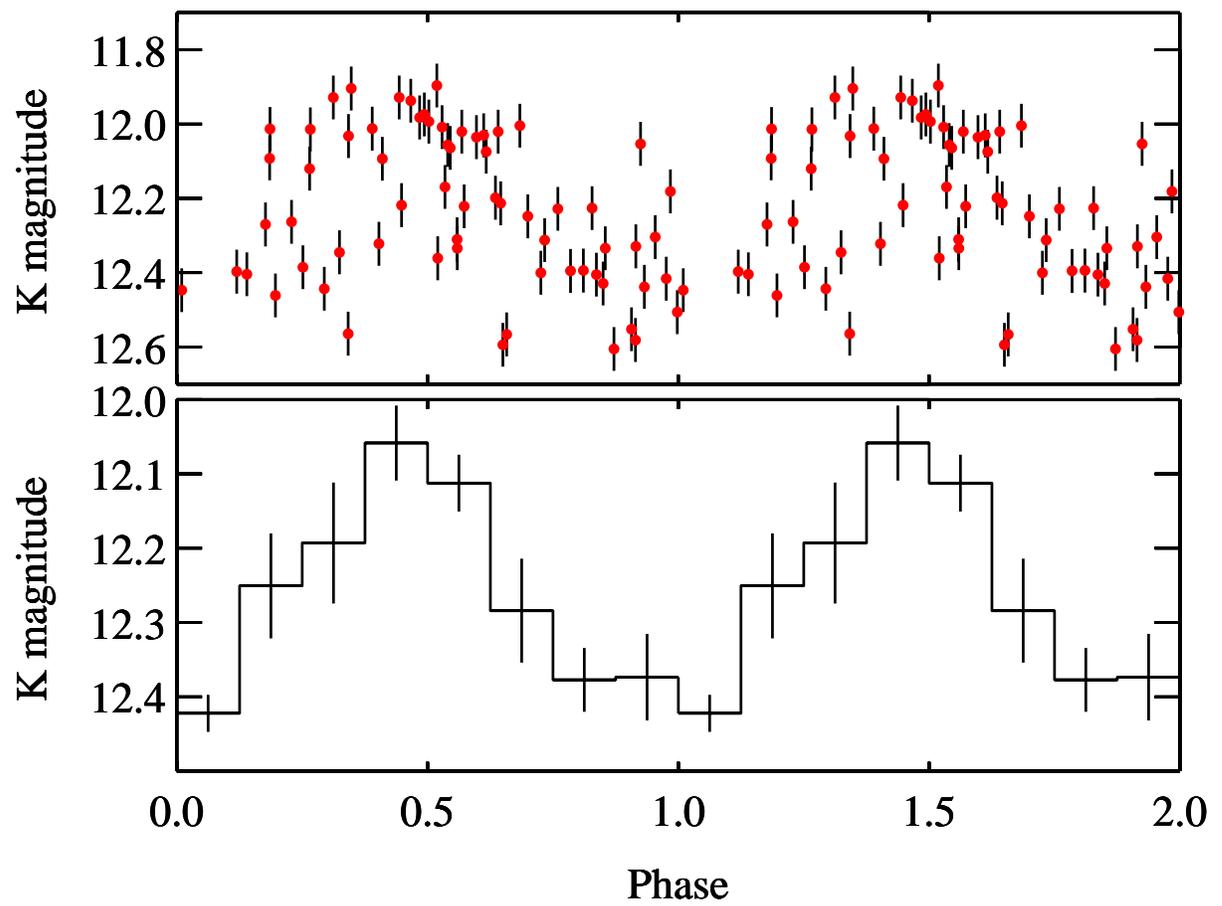}
\figcaption[f8.eps]{
SMARTS CTIO ANDICAM K-band
observations of \src\ folded
on the strongest peak in the power spectrum at 25.8 days.
The two panels show binned and unbinned versions of the
light curve.
\label{fig:ir_fold}
}
\end{figure}

\end{document}